# The Giant Low Surface Brightness Galaxy Malin 1: New Constraints for Its Molecular Gas Mass from GBT/ARGUS Observations

Gaspar Galaz[1], David T. Frayer[2], Matías Blaña[1], J. Christopher Howk[3], Thomas Puzia[1], Evelyn J. Johnston[4], Yasna Ordenes-Briceño[1], Sarah Church[5], Santiago Gil[1], Katerine Joachimi[1], and Marcelo Mora[1]
[1] Instituto de Astrofísica, Pontificia Universidad Católica de Chile, Vicuña Mackenna 4860, Macul, Santiago, Chile; ggalaz@astro.puc.cl
[2] Green Bank Observatory, PO Box 2, Green Bank, WV 24944, USA
[3] Department of Physics and Astronomy, University of Notre Dame, Notre Dame, IN 46556, USA
[4] Núcleo de Astronomía, Facultad de Ingeniería y Ciencias, Universidad Diego Portales, Av. Ejército Libertador 441, Santiago, Chile
[5] Physics Department, Stanford University, 382 Via Pueblo Mall, Satanford, CA 94305, USA
Received 2022 October 18; revised 2022 November 2; accepted 2022 November 8; published 2022 November 28

## Abstract

We report on results from GBT/ARGUS $^{12}$CO(1-0) observations for the giant low surface brightness galaxy Malin 1, which allow us to determine an upper limit for its CO mass, and hence its molecular gas mass and molecular gas mass surface density $\Sigma_{H_2}$. Although we performed very deep observations through 17 hr on-source integration time, reaching a noise level of ~0.2 mK ($T_A^*$) with a corresponding extended source CO limit ($3\sigma$) of 0.09 K km s$^{-1}$, 19 times more sensitive than previous works, we do not detect the $^{12}$CO(1-0) emission line. However, the observations allow us to estimate an upper limit ($3\sigma$) for the CO mass of about $7.4 \times 10^9 \, M_\odot$ for the extended emission, and $1.4 \times 10^8 \, M_\odot$ for the central part of the galaxy. With these figures we conclude that the molecular gas surface density is lower than 0.3 $M_\odot$ pc$^{-2}$, and the corresponding molecular to atomic gas mass ratio is lower than 0.13. The evidence suggests quite different physical conditions for the interstellar medium in Malin 1 compared to that of normal, high surface brightness spirals. This, in one way to another, keeps an usual molecular gas tracer as CO hidden from our observations, in spite of the diverse stellar and structural properties of Malin 1 observed by several authors since more than 30 yr.

*Unified Astronomy Thesaurus concepts:* Galaxies (573); Low surface brightness galaxies (940); Molecular gas (1073); Detection (1911)

## 1. Introduction

Thanks to the pioneering work in the 1970s and 1980s, led by Freeman (1970), Disney (1976), and Sandage & Binggeli (1984), it becomes clear that the universe is populated by a significant fraction of low surface brightness galaxies (LSBGs), some of them with a surface brightness (SB) much fainter than the one measured for the night sky (usually around 22.0 $B$ mag arcsec$^{-2}$). In particular, the discovery by Bothun et al. (1987) of Malin 1, the subject of this Letter, confronts astronomers with a very enigmatic class of LSBG: those with very extended low SB disks, of the size of the Milky Way (MW), or even larger. A typical LSBG has an SB fainter than 23.0 $B$ mag arcsec$^{-2}$. In the last 30 yr, a lot of data has been gathered for LSBGs, and now some clear broad features allow us to distinguish these galaxies from "normal," high surface brightness galaxies (HSBGs). These features can be summarized as follows. It seems that three populations of LSBGs are apparent: (1) One population of relatively small, some irregular galaxies, of the Sm morphological type or dSph type. These galaxies are abundant in the Local Group and are typically satellites of nearby big spirals and elliptical galaxies (Schombert et al. 1995; Grebel et al. 2003; McConnachie 2012; Muñoz et al. 2018). (2) Some very low SB galaxies of the kind described by Sandage & Binggeli (1984), called ultradiffuse galaxies, and rediscovered in large numbers in the Coma cluster by van Dokkum et al. (2015) and Koda et al. (2015) and the Fornax cluster discovered by Muñoz et al. (2015) (see also Eigenthaler et al. 2018; Ordenes-Briceño et al. 2018). These galaxies in general have sizes smaller than or near the size of the MW. Even though some of them are nucleated (Eigenthaler et al. 2018), most look like uniform, very diffuse galaxies. (3) Very large (e.g., giant) and diffuse disk/spiral galaxies, of the size or larger than the MW. Most of them exhibit very faint and extended disks but apparently luminous bulges. In this third category we can include Malin 1, NGC6822, and other giant LSB spiral galaxies. This last category is by far the most intriguing; some of them having disks diameters of more than 80 kpc, evolved stellar populations, and some very diffuse spiral arms. It is worth mentioning that disk/spiral LSBs could have evolved differently than irregular and dwarf LSBs, as have been suggested by Honey et al. (2018). Using Kolmogorov–Smirnov statistical tests on their H I mass, stellar mass, and number of neighbors, they show that the spirals are a statistically different population from the dwarfs and irregulars.

In spite of the impressive different scale lengths of these three types of LSBs, all of them share a common aspect, their low stellar density. This last common feature is what prevents the disk from easily forming gravitational instabilities (Toomre 1977), which finally leads to low stellar formation rates.

Among these giant LSBGs, Malin 1 is perhaps the most striking one. Discovered serendipitously by Bothun et al. (1987), this galaxy shows the most extreme features found in LSBGs, yet it is located at a redshift of $z = 0.082$, i.e., at a distance of 366 Mpc, assuming an accelerated universe, $H_0 = 67.8$ km s$^{-1}$ Mpc$^{-1}$, $\Omega_M = 0.308$, $\Omega_\Lambda = 0.692$. Compared to other giant LSBGs, Malin 1 is the largest one, with also the largest H I mass (Matthews & Gao 2001). It is on its own, one of the largest galaxies detected so far, with a diameter of almost 160 kpc, i.e., 6.5 times the diameter of the MW (Galaz et al. 2015; Boissier et al. 2016). Despite the fact that Malin 1 was discovered more than 30 yr ago, the shape and physical size of its stellar disk and spiral arms were apparent only in 2015, the latter embedded into the faint disk at about 28 $B$ mag arcsec$^{-2}$ (Galaz et al. 2015; Boissier et al. 2016). The H I content for this galaxy is also striking. With a H I mass of $\sim 6.8 \times 10^{10} h_{75}^{-2} M_\odot$ (Pickering et al. 1997), Malin 1 is one of the galaxies with the largest H I mass (and gas fraction) observed so far. Also, the H I disk is as extended as the optical galaxy size from deep optical observations (Lelli et al. 2010), i.e., the H I extension is $\sim 200$ kpc diameter.

Yet, the most enigmatic feature of Malin 1 concerns the amount and physical conditions of its molecular gas. Remarkably, no research has shown the presence of molecular gas in this galaxy so far. In particular, no one has detected carbon monoxide or other $H_2$ typical tracers usually present in spiral galaxies all around the universe. Among the authors trying to detect molecular gas tracers in Malin 1 and in other similar large spiral LSBGs, we can mention the early work by Impey & Bothun (1989), Radford (1992), Knezek (1993), as well as O'Neil et al. (2000), Braine et al. (2000), Moore & Parker (2006), and Das et al. (2006). In particular, Impey & Bothun (1989) and Radford (1992) were the first to adventure for a CO detection in Malin 1. As incredible as it sounds, this first group of authors mistuned by 700 MHz the correct rest frequency of the CO transition at the redshift of Malin 1, and Radford (1992) pointed incorrectly at the center of the galaxy. No one can blame the pitfalls of these authors. Experience shows that the sole approach to understand any of the physical features of Malin 1 seems always risky. Braine et al. (2000) finally pointed at the center of Malin 1 and at the correct frequency, and therefore it can be considered the most precise and pitfall-free effort to detect molecular gas in Malin 1. Although they failed in the detection, their work imposed an interesting upper limit for the CO mass in the galaxy through the $^{12}$CO(1-0) and $^{12}$CO(2-1) transitions, and after reaching a $\sim 2$ mK at a $3\sigma$ significance level.

This led these authors to suggest that perhaps the molecular gas is colder than expected, which in turn reduces the emission per mass of $H_2$, regardless of the fact that the metallicity of Malin 1 is near solar. Although not with Malin 1 as the target, another work worth mentioning is that of Das et al. (2006). In this work, authors observe CO(1-0) and CO(2-1) with BIMA, targeting three giant LSBGs: UGC 5709, UGC 6614, and Malin 2, and detect molecular gas in UGC 6614 and Malin 2. Das et al. (2010) also detected extended molecular gas in Malin 2 using HERA/IRAM 30 m. The CO emission in these galaxies is always very weak, rounding 5 mK, and the estimated molecular masses is between $2 \times 10^8 M_\odot$ and $2 \times 10^9 M_\odot$. It is worth noting that Malin 2, although morphologically similar to Malin 1, is smaller than Malin 1, and also has a brighter SB. Its molecular gas surface density is $\sim 1 M_\odot$ pc$^{-2}$ (Das et al. 2010).

It is important to emphasize that there are indications of stellar formation in some clumps in Malin 1 (Galaz et al. 2015; Boissier et al. 2016, 2018; Junais et al. 2020; Saha et al. 2021), so molecular gas should be present. Although this could make failed searches of molecular gas using CO even more striking, this also could be a clue, since all evidence show H$\alpha$ and other gas phase emission, which is commonly indicative of an already warm interstellar medium (ISM), and eventually in between, cold, shielded molecular gas. We discuss possible scenarios in Sections 4 and 5.

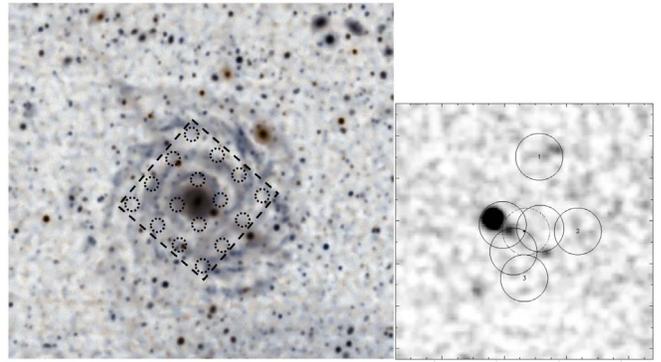

**Figure 1.** Left: ARGUS field superimposed to an optical image of Malin 1 from Galaz et al. (2015). We observed especially the center of Malin 1 and some regions also observed by Braine et al. (2000). The true orientation of the beams will depend on the hour angle of the observations, but we probed the spiral arms of the galaxy too (Galaz et al. 2015; Boissier et al. 2016). Malin 1 is $\sim 3'$ diameter, and the ARGUS footprint is $1'.52 \times 1'.52$, with $30''.4$ spacing between individual elements, each with $7''$ FOV. Right: figure by Braine et al. (2000) at the same scale of the left panel, showing the regions where they pointed the IRAM 30 m telescope to find $^{12}$CO(1-0) and $^{12}$CO(2-1). In Braine et al. (2000) the resolution of the $^{12}$CO(1-0) observations was $22''$, and in our case was $7''$. The lack of a precise picture of the structure of Malin 1, as revealed by this figure, prevented these authors pointing the beam to regions where the molecular gas is more likely present, except the center of the galaxy, clearly visible in the optical image at that time. Compare this right image with the left one, where we have in full view the optical spiral structure of Malin 1. In spite of the better spatial resolution in our observations, we also failed to detect $^{12}$CO(1-0), in the same regions depicted by this figure, as well as others shown in the left panel.

In this Letter we present $^{12}$CO(1-0) observations for Malin 1, using the Byrd 100 m Green Bank telescope with ARGUS, a 16 horn submillimeter detector. Although we failed to obtain a significant detection, we were able to put stringent constraints to the molecular gas amounts. The Letter is organized as follows. In Section 2 we present the technical aspects of the observations and data reduction, in Section 3 we present main results, and discussion in Section 4. Finally, we conclude in Section 5.

## 2. Observations and Data Reduction

Observations were performed with ARGUS on the GBT 100 m dish under two projects (GBT18A-291 and GBT19A-326); the first one was carried out in 2018 December, and the second one ran from 2019 April to November, totaling 17 hr of on-source integration time, over nine observing sessions and 40 hr of total telescope time. ARGUS is a 16-pixel focal plane array designed to operate in the 74-116 GHz range, with a nominal receiver noise of 40–80 K (Sieth et al. 2014). The spatial resolution is about $1.2\lambda/100$ m (Frayer et al. 2019), which yields a measured spatial resolution of about $6''$–$7''$ at 110 GHz. The orientation of the array is $4 \times 4$ elements with beam separations of $30''.4$ on the sky in both the elevation and cross-elevation directions between the elements, with an array field-of-view (FOV) of $1'.52 \times 1'.52$ square (see Figure 1). The polarization is a single linear polarization per feed. The instantaneous instrument bandwidth is 1.5 GHz, and all 16 beams of ARGUS can be connected to the VEGAS 16-channel spectrometer.

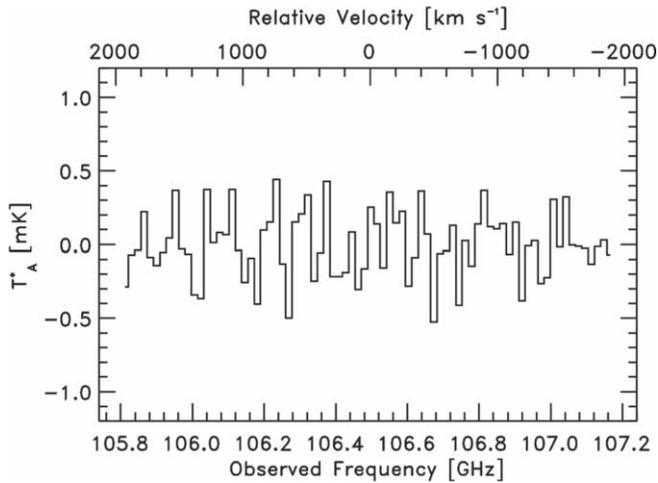

**Figure 2.** Spectrum with the $^{12}$CO(1-0) signal at the central position of Malin 1 with the GBT 7″0 beam. There is no significant emission up to a threshold of 0.23 mK ($T^*_A$, 1$\sigma$). The velocity resolution was smoothed to 50 km s$^{-1}$, and the temperature scale shown is $T^*_A$ (see text for details).

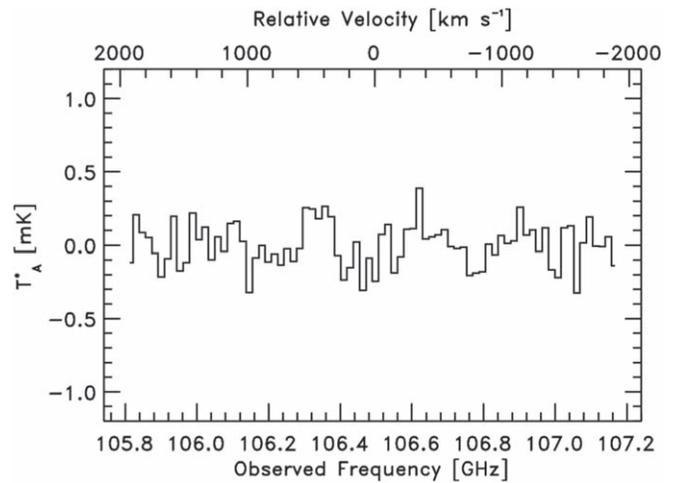

**Figure 3.** Spectrum with the $^{12}$CO(1-0) signal over the 1′.52 square FOV of the ARGUS instrument, where all 16 beams of ARGUS were combined. There is no significant emission up to a threshold of 0.15 mK ($T^*_A$, 1$\sigma$). The velocity resolution was smoothed to 50 km s$^{-1}$, and the temperature scale shown is $T^*_A$ (see text for details).

Observations were carried out at the observed redshifted frequency of $^{12}$CO(1-0) of 106.4861 GHz. The initial six observing sessions consisted of NOD observations between two beams within the array separated by 60″8 on the sky. The instrument FOV with the individual elements with their relative position on the galaxy is shown in Figure 1. The telescope was nodded to alternate the source between the two beams with 60 s scans. The raw spectral resolution was 4.1 km s$^{-1}$ per channel with 1024 channels. The effective spatial resolution is 7″, equivalent to 11 kpc at the distance of the source ($z = 0.0825$), and the NOD observing technique would be sensitive to detecting emission on spatial scales of less than 1′. After not detecting CO emission on these spatial scales, the last three observing sessions were done in the ON–OFF mode to look for extended emission that may have been subtracted out by the NOD observations. The OFF position was well outside the FOV of the instrument, and the data for all 16 beams were coadded to provide sensitivity of CO emission on spatial scales of $\sim$2′.

We use the GBTIDL reduction package for GBT to carry out the data reduction. The data were calibrated to the $T^*_A$ temperature scale following the standard calibration methodology for ARGUS (Frayer et al. 2019). At the observed frequency, the conversion between main-beam temperature and $T^*_A$ is $T_{mb}/T^*_A = 2.9$ (Frayer et al. 2019). The most critical aspect of the reduction is the removal of data with spurious baseline structures that can dominate the coadds and obfuscate an astronomical signal. Manual inspections were made to subsets of data to derive noise threshold parameters associated with bad data and to objectively remove noisy baseline data from the coadds. By removing approximately 20% of the data, the resulting data were of high-quality showing baselines that were flat, and the noise properties of the data integrated down as expected theoretically with increasing time. Each spectrum had a low-order polynomial removed across the full bandwidth (4000 km s$^{-1}$), and good data were coadded together and then smoothed to 50 km s$^{-1}$ to produce the final spectra. The 1$\sigma$ limit for the NOD data (for spatial scales similar to the beam size up to 1′ in size) is 0.23 mK ($T^*_A$, see Figure 2), while the 1$\sigma$ limit for the ON–OFF data (for spatial scales of $\gtrsim$2′) is 0.15 mK ($T^*_A$, see Figure 3).

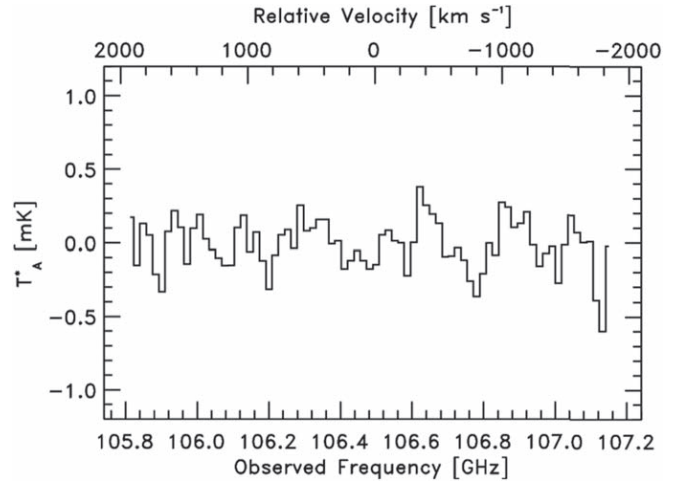

**Figure 4.** Same as Figure 3, but correcting from disk peculiar velocity induced by disk rotation. The correction is made using the projection of the rotation curve from Lelli et al. (2010), which is derived using the H I data of Pickering et al. (1997) and assuming that the molecular gas has the same rotation profile.

### 3. Results

Figure 3 shows the reduced, summed, and final spectrum (in $T^*_A$), where the $^{12}$CO(1-0) emission line is expected. No indication for a line emission is evident, with a 1$\sigma$ noise level of $\sim$0.15 mK, with a channel resolution smoothing of 50 km s$^{-1}$.

Figure 4 shows the same results as in Figure 3, but where we subtracted for each individual spectra the projected local rotation velocity of the H I disk before combining them. For this we used the rotation curve of Lelli et al. (2010) who derived it analyzing the H I emission observations of Pickering et al. (1997). We assume that the molecular gas is acting under the same rotation profile. Although we average radially the rotation curve from Pickering, the major change of velocity is as a function of radii, not the azimuth. Figure 4 does not show a significant difference than Figure 3, indicating thus that the velocity corrections do not change conclusions: we do not detect CO emission with a $1\sigma$ noise level of $\sim 0.15$ mK, with a channel resolution smoothing of 50 km s$^{-1}$.

Nevertheless, this nondetection is useful to provide an upper limit to the CO mass of Malin 1, and then the H$_2$ mass upper limit. In other words, these observations secure that Malin 1 has a molecular gas mass no larger than a certain value. This imposes useful and realistic constraints of the molecular gas mass for this giant and very diffuse galaxy. And, as it will be explained in what follows, we reach an *even lower* upper limit for the molecular mass for Malin 1 compared to previous works. It is worth noting that our resolution of $\sim 7''$, equivalent to 11 kpc at the distance of Malin 1, does not allow us to resolve individual giant molecular clouds, which have sizes of $\sim 50$ pc. However, the clumps observed in Malin 1 (Galaz et al. 2015; Boissier et al. 2016), could nest giant molecular clouds associations. These clumps were included in the the work of Braine et al. (2000), with no positive results (see Figure 1). In our case, we map at least two of the clumps, without success.

### 3.1. An Upper Limit for the Molecular Gas Mass: Comparison with Previous Works

To place this work in context, Braine et al. (2000) using IRAM 30 m submillimeter telescope, obtained no $^{12}$CO(1-0) detection corresponding to a $3\sigma$ molecular mass limit of $2.7 \times 10^9 M_\odot$, adopting a distance[6] of 366 Mpc for Malin 1. In order to compare our work to that of Braine et al. (2000), we adopt the same methodology and physical values for the source.

We estimate the *upper limit* for the H$_2$ mass in the same way as Braine et al. (2000) did, using

$$M_{H_2}^{up} = I_{CO}^{lim} \times 2m_p \times [N(H_2)/I_{CO}] \times \Omega D^2, \quad (1)$$

where $I_{CO}^{lim}$ is the measured CO line limit in K km s$^{-1}$, $m_p$ is the mass of a proton, $\Omega$ is the beam size in steradians (in this case a Gaussian beam size of $7''$ yields an area over the sky of $1.305 \times 10^{-9}$ str), and $D$ is the distance to the source (366 Mpc).

We adopt the standard estimated Galactic conversion factor $N(H_2)/I_{CO(1-0)} = 2 \times 10^{20}$ cm$^{-2}$ (K km s$^{-1}$)$^{-1}$ (Bolatto et al. 2013), the same used by Braine et al. (2000). $I_{CO}^{lim}$ is computed using the relation

$$I_{CO}^{lim} = \sigma \sqrt{\Delta V \delta v}, \quad (2)$$

where $\sigma$ is the $T_{mb}$ rms noise level of 0.67 mK ($1\sigma$) for the NOD data of the central core shown in Figure 2, $\Delta V$ is the adopted line width of 320 km s$^{-1}$ based on the H I line width at the central core (Braine et al. 2000), and $\delta v$ is the channel width of 50 km s$^{-1}$. This yields $I_{CO}^{lim} = 0.085$ K km s$^{-1}$ ($1\sigma$). Using Equation (1), this corresponds to a $3\sigma$ upper limit to the molecular gas mass (not including He) of $M_{H_2} < 1.4 \times 10^8 M_\odot$ distributed over a central area encompassing $7''$ Gaussian beam size.

This H$_2$ mass upper limit is 19 times more sensitive than the previous one reported by Braine et al. (2000), and represents a limit on the molecular gas in the central core of the galaxy, which could be reasonably expected for normal galaxies. In terms of sensitivity limit for the core, the 19 times improvement of the GBT over the 30 m is based on the emission being smaller than the beam size (point-source limit). If the molecular gas is contained in small clumps throughout the optical extent of the galaxy, then the ON–OFF data that are sensitive to emission on a large scale, would provide a more applicable constraint. Using the $3\sigma$ noise level associated with the extended emission of 1.3 mK ($T_{mb}$, 50 km s$^{-1}$), and assuming a line width of 100 km s$^{-1}$ for the outer regions as seen in H I (Braine et al. 2000), the corresponding extended source CO line limit ($3\sigma$) is $I_{CO}^{lim} < 0.09$ K km s$^{-1}$. Using Equation (1) to constrain the extended molecular gas, the value of $\Omega$ of the observations corresponds to the footprint of the ARGUS array of $1.52' \times 1.52'$ over which the measurements were averaged. This results in an extended molecular gas limit ($3\sigma$) of $7.4 \times 10^9 M_\odot$ (averaged over $1'.52 \times 1'.52$ of Malin 1).

### 4. Discussion

With the figures obtained in the precedent section, we can estimate an upper limit for the molecular gas surface density ($\Sigma_{H_2}$ hereafter), an important parameter that indirectly defines the stellar formation efficiency (Krumholz et al. 2009; Wong et al. 2013). Assuming a H$_2$ mass upper limit of $7.4 \times 10^9 M_\odot$ distributed over $1.52' \times 1.52'$, a distance[7] to Malin 1 of 366 Mpc, we obtain an upper limit $\Sigma_{H_2} < 0.27 M_\odot$ pc$^{-2}$. Using the own data and parameters from Braine et al. (2000), we compute a $\Sigma_{H_2} < 0.2 M_\odot$ pc$^{-2}$. Thus, our value is slightly higher than that from these authors. On the other hand, using the H I measures for Malin 1 by Pickering et al. (1997), we obtain a H I mass surface density of $\Sigma_{HI} = 2.3 M_\odot$ pc$^{-2}$. These H$_2$ and H I mass surface densities allow us to compute for Malin 1 an *upper limit* for the molecular to atomic gas mass ratio, yielding $M_{H_2}/M_{HI} = 0.12$. The corresponding value from Braine et al. (2000), transformed to an accelerated universe, gives $M_{H_2}/M_{HI} = 0.09$.

Our upper $M_{H_2}/M_{HI} = 0.12$ ratio is higher than some values presented in other works for LSBGs. In particular, it is 1.3 (30%) times larger than the one obtained for Malin 1 by Braine et al. (2000). We claim that this difference, however, is not significant. On the other hand O'Neil et al. (2000), for the first LSB galaxy for which a CO detection was successful, report a ratio of 0.15, similar to our corresponding upper limit of 0.13. It is worth noting that the excellent discussion by O'Neil et al. (2000) on the "observed" versus "the Mihos et al. (1999) predicted" values for $M_{H_2}/M_{HI}$ is also applicable here. In agreement with O'Neil et al. (2000), but also in disagreement with Mihos et al. (1999) theoretical values for generic LSBGs, we are finding a lower $M_{H_2}/M_{HI}$ ratio for Malin 1. At worst, considering that our value is an upper limit, the actual ratio could be even lower.

Our value for the upper limit of the molecular gas mass surface density $\Sigma_{H_2} < 0.27 M_\odot$ pc$^{-2}$) is clearly lower than all corresponding values for the sample of infrared (IR) bright galaxies from Wong et al. (2013), where the average value is $\Sigma_{H_2} \sim 7$–$10 M_\odot$ pc$^{-2}$ and the lower value they measure is $\Sigma_{H_2} \sim 0.8 M_\odot$ pc$^{-2}$. Thus, we are obtaining for Malin 1 an upper value for $\Sigma_{H_2}$ much lower than those obtained for IR luminous galaxies, in agreement with the fact that Malin 1 lacks IR emission so far (Rahman et al. 2007), but larger than those obtained in other works for LSBGs (Matthews &

---

[6] Note that Braine et al. (2000) adopts a distance to Malin 1 of 250 Mpc. Here we transformed their H$_2$ upper mass limit to a distance to Malin 1 of 366 Mpc.

[7] Note that $M_{H_2}^{lim}$ depends on $D$, the distance to the source, and therefore it is sensitive to cosmology. Values for Malin 1 from Braine et al. (2000) must be multiplied by $(366/250)^2 = 2.14$.

Gao 2001). Note however that the figures derived here are upper limits, and therefore actual values for these quantities for Malin 1 could be even lower.

It is worth noting that the H I mass surface density $\Sigma_{HI} = 2.3 \, M_\odot \, \text{pc}^{-2}$ of Malin 1 is also lower than the value normally found in luminous IR galaxies. From the same paper (Wong et al. 2013), we conclude that the typical $\Sigma_{HI}$ is $\sim 7-10 \, M_\odot \, \text{pc}^{-2}$, $\sim 4$ times higher than the value for Malin 1. This seems to be the situation for many LSB galaxies measured by Pickering et al. (1997) in their excellent work, where they find lower H I mass surface densities for LSBGs, compared to those obtained for high SB spirals. The scenario of galaxies with smaller molecular to atomic gas ratios matches with the one suggested by Mihos et al. (1999): LSB galaxies tend to have larger fractions of H I with respect to H$_2$ due to the low conversion of H I in H$_2$, this is caused in turn by the low amount of grains, which seems to be the case for Malin 1.

This Letter agrees with conclusions of Braine et al. (2000), who argue that the disk of Malin 1 is clearly well below the luminosity of a spiral's disk of similar stellar formation rate. We emphasize that the conversion factor, at the scale we observed here, is probably similar to that acting in other normal spirals. Our results clearly show that when located at the Kennicutt–Schmidt diagram (Kennicutt 1998), Malin 1 is at the extreme left (and lower) side of the diagram, at the limit of the star formation efficiency per molecular gas mass. It is also interesting to note that although most of the disk of Malin 1 seems to be under the critical density to allow instability to form stars, there is a region near 50 kpc from the center where the H I mass density is higher the critical H I mass density (Pickering et al. 1997). The regions where the stellar formation is suspected (Galaz et al. 2015) show no emission.

Regarding the reasoning of Braine et al. (2000), who argue that the $^{12}$CO(1-0) is not detected because it could be too cold, limiting the efficiency of the collision induced CO radiation, and that a higher frequency radiation of higher order, like $^{12}$CO(3-2), is even more difficult to be detected, we have a different hypothesis. The H$_2$ could be *warmer* than that in a HSBG, as suggested in the past by Mihos et al. (1999). These authors argue that in an LSBG, less than 20% of the molecular gas is at temperatures colder than 30 K. This has sense, since it is expected that in a quite transparent ISM like the one prevalent in Malin 1, UV radiation is able to reach larger distances. In particular, the UV radiation from the LINER activity in the center of Malin 1 could reach the disk and prevent the gas from forming stars (Barth 2007). Therefore, in the case of this galaxy with a low density disk, the diluted and low density ISM can be also quite warm due to the penetration of the UV light. In the case of Malin 1, there is some emission in the UV detected with GALEX, as shown by Boissier et al. (2016), indicating that clumps of young stars are indeed present, especially in the regions R1, R2, and R3 depicted in Figure 3 of Galaz et al. (2015).

Another clue to consider is that the disks in LSBGs appear similar to extended UV disks (Thilker et al. 2007, XUV): they seem to trace last episodes of evolution of the galaxy. These disks show star formation in the UV in clumps, similar to Malin 1 (Boissier et al. 2016), and in only a few of them molecular gas has been detected via CO, such as the case of M63 (Dessauges-Zavadsky et al. 2014) and M83 (Koda et al. 2022), but in very compact clumps. Should CO be present in Malin 1, it would probably be located in such clumps, and could be the reason why we do not detect CO with the GBT observations.

## 5. Conclusions

In this work we presented submillimeter observations from the Byrd 100 m single dish Green Bank Telescope to detect $^{12}$CO(1-0) emission in Malin 1. This is another effort along many others executed in the last 30 yr to detect molecular gas in this giant low surface brightness spiral galaxy. As in previous efforts, our detections were negative, in spite of reaching a CO line limit at $3\sigma \, I_{CO}^{lim} < 0.09$ K km s$^{-1}$. This low detection threshold suggests an upper limit for the CO mass of $M_{H_2} < 1.4 \times 10^8 \, M_\odot$ ($3\sigma$) in the center, 19 times lower than the corresponding upper limit derived by previous works (Braine et al. 2000), and a value of $M_{H_2} < 7.4 \times 10^9 \, M_\odot$ ($3\sigma$) for the extended emission in the galaxy, over $1'.52 \times 1'.52$. This upper limit mass implies a molecular gas mass surface density $\Sigma_{H_2} < 0.27 \, M_\odot \, \text{pc}^{-2}$, clearly at the lower end ($< 3 \, M_\odot \, \text{pc}^{-2}$) of the Kennicutt–Schmidt law (Kennicutt 1998; Krumholz et al. 2009; Wyder et al. 2009). This last figure, along with the available H I mass from other works, allows us in turn to derive an upper limit for the molecular to atomic gas mass ratio for Malin 1, which gives $M_{H_2}/M_{HI} < 0.12$.

After all the past and present evidence about the rich stellar content of Malin 1 distributed over a radius of up to $\sim 160$ kpc (Barth 2007; Galaz et al. 2015; Boissier et al. 2016), and the many negative efforts to detect molecular gas emission in this giant LSB galaxy (Braine et al. 2000, and references therein), the questions about where the molecular gas of Malin 1 is, and under what physical conditions it behaves, are even more challenging.

Among the plausible clues to answer the above questions, we can mention (1) the conversion factor in Malin 1 is significantly different (higher) to that obtained in our galaxy; (2) the ISM is warmer than expected and finally the CO emission in the lower $J$ transitions is finally not significant. This hypothesis is in agreement with the predictions by Mihos et al. (1999), who predict that the ISM in LSB galaxies should be warmer, at temperatures above 30 K, especially for an homogeneous ISM. In this situation, the stellar formation is only suitable in clumps; the only places where the shielding secure lower ISM temperatures; (3) by some reason, the CO emission is completely shifted in velocity and we are not detecting it by tuning the receivers at the galaxy redshift. Adding more complexity to the already perplexing situation of not detecting emission from CO in Malin 1, evidence from Galaz et al. (2015) and Boissier et al. (2016) suggests that clumps of gas/stellar formation activity is present in many places in the galaxy, in agreement with the analysis of Braine et al. (2000). Why is there no CO emission in these places, where we even see UV emission from GALEX (Boissier et al. 2016)? It is a quite disturbing aspect of the elusive molecular gas emission of Malin 1 that nor the center, nor these blobs of apparent molecular gas reservoirs in the disk (see Figure 1) exhibit CO emission to the very low detection thresholds probed here. New observations should invest more time in these locations individually.

Finally, it could be also a good strategy to use other molecular gas tracers better suited for the ISM conditions that likely exist in a galaxy like Malin 1. These lines, which can be devised for example using a model like RADEX (van der Tak et al. 2007), are probably those more efficiently produced in a low density but warmer molecular gas (see for example Galaz et al. 2008, who used $^{12}$CO(3-2) to detect CO in a reduced set of LSB galaxies). Perhaps these lines are better suited to detect molecular gas tracers than the lines associated to CO collisional transitions. For example, the [CI] line could be one of them. This line might help to detect possible dark molecular clouds hidden in an ISM condition such as in Malin 1, either by finding strong [CI] emission and weak/absent CO emission or looking for enhanced emission regions of [C/CO] abundance. These regions are expected in the outer parts of molecular clouds, the well-known photodissociation regions (Burton et al. 1990, 2015).

Authors acknowledge the access to ARGUS at the Byrd 100 m telescope at the Green Bank Observatory. Green Bank Observatory is supported by the National Science Foundation and is operated by Associated Universities, Inc. G.G. acknowledges the excellent support of the GBT staff during a training internship, as well as the assistance during remote observations from Chile. G.G., M.B., T.P., S.G., K.J., Y.O., and M.M. gratefully acknowledge support by the ANID BASAL projects ACE210002 and FB210003. E.J.J. acknowledges support from FONDECYT Iniciación en investigación 2020 Project 11200263. Y.O. acknowledges the support of Fondecyt Postdoctorado 3210442. We acknowledge the suggestions made by an anonymous referee, which improved this Letter.

This research has made use of NASA's Astrophysics Data System Bibliographic Services. This research has made use of the NASA/IPAC Extragalactic Database (NED), which is operated by the Jet Propulsion Laboratory, California Institute of Technology, under contract with the National Aeronautics and Space Administration.

### ORCID iDs

Gaspar Galaz https://orcid.org/0000-0002-8835-0739
David T. Frayer https://orcid.org/0000-0003-1924-1122
Matías Blaña https://orcid.org/0000-0003-2139-0944
J. Christopher Howk https://orcid.org/0000-0002-2591-3792
Thomas Puzia https://orcid.org/0000-0003-0350-7061
Evelyn J. Johnston https://orcid.org/0000-0002-2368-6469
Yasna Ordenes-Briceño https://orcid.org/0000-0001-7966-7606
Santiago Gil https://orcid.org/0000-0001-5551-3872
Katerine Joachimi https://orcid.org/0000-0001-5348-7543
Marcelo Mora https://orcid.org/0000-0003-4385-0411